\documentclass[letterpaper,aps,pra,showpacs,twocolumn]{revtex4}
\usepackage{amsmath}
\usepackage{graphicx}
\usepackage{float}
\usepackage[ansinew]{inputenc}
\usepackage{array}
\usepackage{color}
\usepackage{amsxtra}
\usepackage{amstext}
\usepackage{amssymb}
\usepackage{latexsym}
\usepackage{gensymb}
\usepackage{dsfont}
\usepackage{braket}
\usepackage{color}
\begin{document}

%\preprint{}

\title{Single and two-mode mechanical squeezing of an optically levitated nanodiamond via dressed-state coherence}


% \affiliation can be followed by


\author{Wenchao Ge}
\author{M. Bhattacharya}
\affiliation{School of Physics and Astronomy, Rochester Institute of Technology, 84 Lomb Memorial Drive,
Rochester, NY 14623, USA}

\date{\today}
%\date{\today}
\begin{abstract}
Nonclassical states of macroscopic objects are promising for ultrasensitive metrology as well as testing quantum
mechanics. In this work, we investigate dissipative mechanical quantum state engineering in an optically levitated
nanodiamond. First, we study single-mode mechanical squeezed states by magnetically coupling the mechanical motion to a dressed three-level 
system provided by a Nitrogen-vacancy center in the nanoparticle. Quantum coherence between the dressed levels is created via microwave fields to induce a two-phonon transition, which results in mechanical squeezing. Remarkably, we find that in ultrahigh vacuum 
quantum squeezing is achievable at room temperature with feedback cooling. For moderate vacuum, quantum squeezing is possible with cryogenic temperature. Second, we present a setup for two mechanical modes coupled to the dressed three levels, which results in two-mode squeezing analogous to the mechanism of the single-mode case. In contrast to previous works, our study provides a deterministic method for engineering macroscopic squeezed states without the requirement for a cavity.
\end{abstract}
%\keywords{}

\maketitle
%more references on microscopic level

\section{Introduction}
Optical levitation has been a powerful tool for trapping and manipulating small particles since its inception 
\cite{Ashkin:70}. Recent advances with optically levitated dielectric microscopic and nanoscopic particles have 
provided a promising platform for optomechanics \cite{Aspelmeyer:14} with multiple degrees of freedom and ultrahigh 
mechanical quality factors \cite{Chang:10, Yin:13, Neukirch:15, Shi:15}. Motivated by testing quantum mechanics 
at the macroscopic scale and by potential applications in nanoscale sensing, many studies have been performed 
on the center of mass motion cooling \cite{Li:11,Gieseler:12, Arita:13,Kiesel:13,Genoni:15,Rodenburg:16, Jain:16}, 
quantum state preparation \cite{Romero:11}, non-equilibrium dynamics \cite{MillenJ.:14,Gieseler:14nn,Gieseler:15,Ge:16}, 
and ultra-sensitive metrology \cite{Geraci:10,Arvanitaki:13,Moore:14,Ranjit:15} of optically trapped nanoparticles.

Recently, levitated nanoparticles with internal degrees of freedom, such as the nitrogen-vacancy (NV) center with a 
single spin, have been studied theoretically to test quantum wavefunction collapse models \cite{Yin:13pra, Scala:13} 
and quantum gravity \cite{Albrecht:14} in vacuum. More recently, optical levitation of nanodiamonds in low vacuum 
has been demonstrated experimentally \cite{Neukrich:15np, Hoang:15}, paving the way for preparing quantum states of 
mechanically oscillating levitated nanoparticles.

In this article, we propose a method for creating single- and two-mode squeezed states of mechanical oscillation
of an optically levitated single NV center nanodiamond, motivated by the potential for the applications of such states
to sensitive metrology \cite{Loudon:87}. Generally, single-mode mechanical squeezing has been proposed theoretically
\cite{Jaehne:09,Nunnenkamp:10, Liao:11, Kronwald:13, Gu:13, Didier:14,Lv:15, Wang:16,Lotfipour:16, Agarwal:16} and
demonstrated experimentally \cite{Pirkkalainen:15, Lecocq:15, Wollman:15, Lei:16} in cavity-based optomechanical systems,
for example, by driving an optomechanical cavity with two  frequency tones
\cite{Kronwald:13, Pirkkalainen:15, Lecocq:15, Wollman:15, Lei:16}. Also, two-mode mechanical squeezing has been studied
via mechanisms such as dissipative reservoir engineering \cite{Tan:13,Woolley:14}, quantum measurement backaction
\cite{Nielsen:16}, and nondegenerate parametric amplification \cite{Patil:15, Cheung:16}.
 %squeezed light \cite{Jaehne:09}, via quadratic coupling \cite{Nunnenkamp:10}, by modulating the driving field
 %\cite{Liao:11}, via reservoir engineering dissipative mechanism \cite{Kronwald:13, Gu:13, Didier:14}, via a nonlinear
 %duffing term \cite{lv:15}, by coupling to an auxiliary cavity \cite{Wang:16}, parametric amplifier \cite{Agarwal:16}.
 %Recently, squeezing of mechanical motion has been demonstrated in microwave cavity optomechanical systems via two-tone driving %\cite{Pirkkalainen:15, Lecocq:15, Wollman:15, Lei:16}.
More specifically, spin-mechanical systems \cite{Yin:15} have been studied extensively, using
strained-induced coupling \cite{Bennett:13, MacQuarrie:13,Ovartchaiyapong:14, Golter:16}, or in the presence of a
magnetic field gradient \cite{Rabl:09, Kolkowitz:12, Arcizet:11}, for mechanical cooling
\cite{Rabl:09,MacQuarrie:16,Zhang:13, Yan:16}, optomechanical spin control \cite{Bennett:13, Golter:16}, and mass
spectrometry \cite{Zhao:14}. Recently, single-mode mechanical squeezing was investigated via qubit measurement 
\cite{Rao:16} and feedback stabilization \cite{Genoni:15a} in a spin-mechanical system.

In the present work, the nanoparticle mechanical motion is coupled to the single NV center spin via a magnetic field gradient,
without requiring a cavity \cite{Scala:13,Yin:13pra}. Distinct from the works cited above \cite{Rao:16,Genoni:15a}, our method does not require a 
measurement-based technique, but instead relies on a microwave field-induced spin-state coherence for generating
steady-state mechanical squeezing in both the single-mode and two-mode cases. By applying two microwave fields coupling 
the $\ket{0}$ and $\ket{\pm1}$ states of the NV center ground-state triplet \cite{Rabl:09}, a dressed three-level system 
is created to induce a two-phonon transition in the mechanical oscillator, an interesting effect which has not been studied 
before, to the best of our knowledge, in spin-optomechanical systems. 

To arrive at our results, we employ a master equation approach to describe the mechanical motion, by tracing out the spin degree of freedom in 
the Born-Markov approximation. This approach is enabled by applying optically-induced dissipation \cite{Doherty:13} 
to the spin triplet states leading to relaxation rates much stronger than the spin-mechanical coupling. For the 
single-mode case, we find remarkably that quantum squeezing is achievable at room temperature with experimentally achievable ultrahigh vacuum and feedback cooling techniques \cite{Jain:16}. For moderate vacuum, quantum squeezing is possible with precooled phonon occupation number. For the two-mode case, we propose a setup such that both modes are coupled to the dressed states in exactly the same way as for the single-mode case. The analytical results for both the single-mode and the two-mode squeezing are equivalent to each other. We also present numerical results in a wide range of parameters for single-mode squeezing, which is applicable to the two-mode case.  

The analysis presented using an optically levitated nanodiamond is quite general, therefore the proposal 
can also be extended to related systems, such as, nanodiamonds using magneto-gravitational traps  \cite{Hsu:16} or Paul traps \cite{Kuhlicke:14,Delord:16} , which avoid optical scattering, or a single NV center coupled to an cantilever \cite{Rabl:09}.

%Q$\sim10^3$ recently realized, we require Q $\sim 10^6$ which is within the reach of future optically levitated

\section{Single NV center coupled to one mechanical mode}
\label{sec:ModelNV1}
\subsection{The model}
\begin{figure}[tbp]
\centering
\includegraphics[width=0.95\columnwidth]{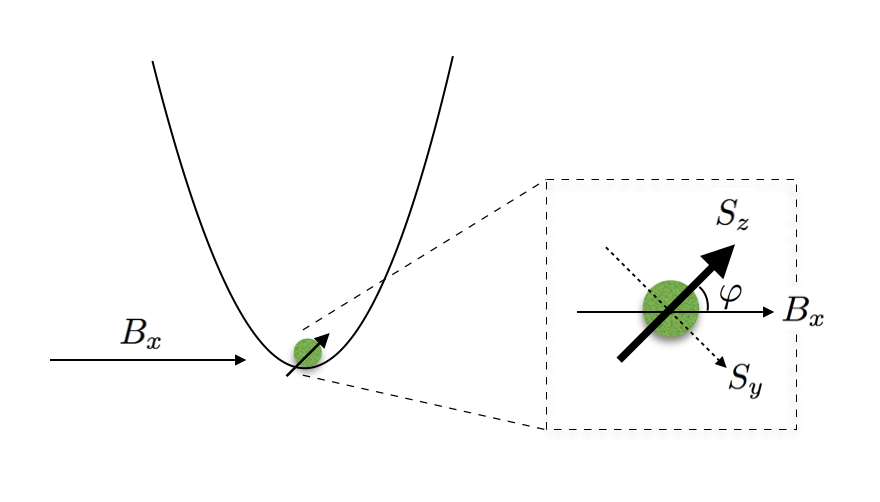}
\caption{The configuration considered in Section \ref{sec:ModelNV1}. The green circle denotes an optically levitated 
nanodiamond oscillating in a harmonic potential (black curve) along the $x$ coordinate. A magnetic field is also applied along the same 
direction. The arrow on the circle denotes the direction of the spin $S_z$ axis corresponding to an NV center contained in the 
nanodiamond. We note that the spin axes are not aligned with the coordinate axes in general. As shown in the figure, $S_{y},S_{z}$
and $B_{x}$ all lie on the same plane. Not shown is $S_{x}$, which is perpendicular to $B_{x}$ and points out of the 
plane of the paper.
\label{fig:Scheme}}
\end{figure}

\begin{figure}[tbp]
\centering
\includegraphics[width=1.0\columnwidth]{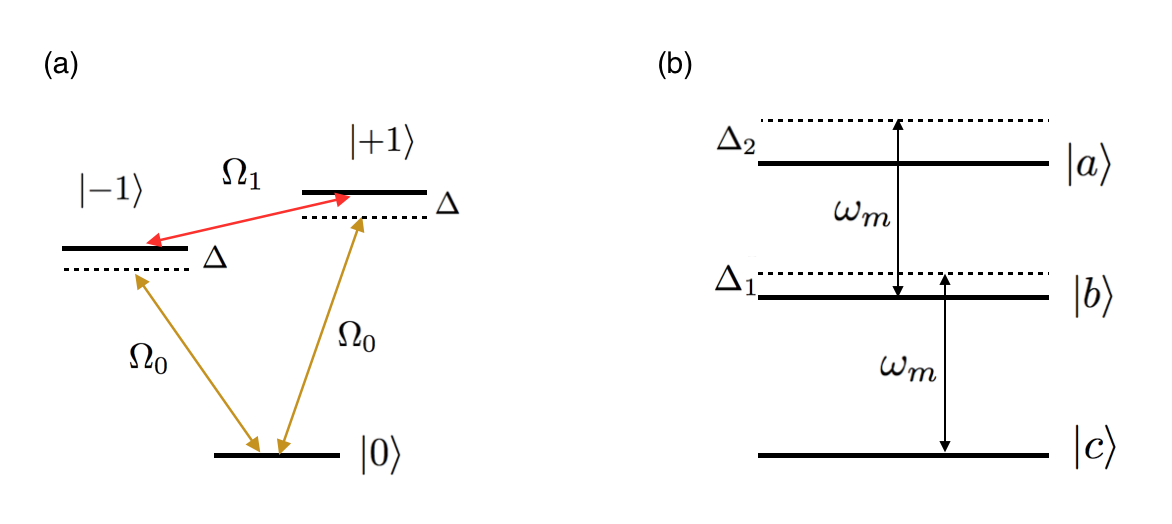}
\caption{(a) The bare energy levels of the NV center $\ket{0}$ and $\ket{\pm 1}$, the Rabi frequencies of the microwave fields
$\Omega_{0}$ and $\Omega_{1}$ and the detuning $\Delta$. (b) The dressed states of the NV center, $\ket{a},\ket{b}$ and $\ket{c}$ 
defined in Eq.~(\ref{eq:DressedStates1}), the oscillator phonon energy $\omega_{m}$ and the effective detunings $\Delta_{1}=\omega_m-\omega_{bc}$ and 
$\Delta_{2}=\omega_m-\omega_{ab}$.
\label{fig:SpinLevels1}}
\end{figure}

We consider a single NV center nanodiamond optically trapped in vacuum and executing harmonic center of mass motion 
along all three directions in space, as shown in Fig.~\ref{fig:Scheme}. 
A magnetic field $B_x=B_0 x$ with the gradient $B_0$ is applied to couple the mechanical motion and the electron 
spin of the NV center. The magnetic field is assumed to be lying in the $z-y$ plane of the spin axes and making an angle $\varphi$ 
with the $z$ axis. The Hamiltonian of the system is
\begin{equation}
\label{eq:BigH}
\begin{split}
\mathcal{H}&=\mathcal{H}_{\text{m}}+\mathcal{H}_{\text{NV}}+\mathcal{H}_{\text{int}},\\
\mathcal{H}_{\text{m}}&=\hbar\omega_m d^{\dagger}d,\\
\mathcal{H}_{\text{NV}}&=-\hbar\Delta\left(\ket{+1}\bra{+1}+\ket{-1}\bra{-1}\right)\\
&+\frac{\hbar\Omega_0}{2}\left(\ket{0}\bra{+1}+\ket{0}\bra{-1}+h.c.\right)\\
&+\frac{\hbar\Omega_1}{2}\left(\ket{-1}\bra{+1}+\ket{+1}\bra{-1}\right),\\
\mathcal{H}_{\text{int}}&=\hbar g\cos(\varphi)S_z(d^{\dagger}+d)+\hbar g\sin(\varphi)S_y(d^{\dagger}+d),
\end{split}
\end{equation}
where $\omega_m$ is the mechanical oscillation frequency determined by the optical trap beam intensity and the nanoparticle
mass $m$, $g=g_l\mu_{B}B_{0} x_{0}/\hbar$, $g_l\approx 2$ is the Land\'{e} factor, $\mu_{B}$ is the Bohr magneton,
$x=x_0(d^{\dagger}+d)$, and $x_0=\sqrt{\hbar/2m\omega_{m}}$. The creation (annihilation) operator of the mechanical
motion along $x-$axis is $d^{\dagger}$ ($d$). The spin operator components are $S_z=\ket{+1}\bra{+1}-\ket{-1}\bra{-1}$, and $S_y=-i\ket{+1}\bra{-1}+i\ket{-1}\bra{+1}$. The NV center Hamiltonian has been obtained in the rotating-wave frame with two
microwave driving frequencies which couple the $\ket{0}$ and $\ket{\pm1}$ states of the spin-1 system with a detuning $\Delta$ and Rabi
frequency $\Omega_0$, as shown in Fig.~\ref{fig:SpinLevels1} (a). 
By going to the eigenbasis of $\mathcal{H}_{\text{NV}}$, we find that
\begin{equation}
\label{eq:DressedH1}
\begin{split}
\mathcal{H}&=\hbar\omega_m d^{\dagger}d+\hbar\omega_{ac}\ket{a}\bra{a}+\hbar\omega_{bc}\ket{b}\bra{b}\\
&+\hbar\left(g_{s}\ket{c}\bra{b}+g_{c}^{\ast}\ket{b}\bra{a}+h.c.\right)(d^{\dagger}+d),
\end{split}
\end{equation}
where the coupling constants $g_{s}=-g e^{i\varphi}\sin(\theta)$, $g_{c}=ge^{i\varphi}\cos(\theta)$, and the dressed
states are
\begin{eqnarray}
\label{eq:DressedStates1}
\ket{a}&=&\sin(\theta)\ket{0}+\cos(\theta)\ket{+},\nonumber\\
\ket{b}&=&\ket{-},\\
\ket{c}&=&\cos(\theta)\ket{0}-\sin(\theta)\ket{+},\nonumber 
\end{eqnarray}
with $\ket{\pm}=(\ket{+1}\pm\ket{-1})/\sqrt{2}$, and $\tan(2\theta)=-\sqrt{2}\Omega_0/(\Delta-\Omega_1/2)$.
The eigenvalues of the dressed states are 
$\omega_{a,c}=\left(-\Delta+\Omega_1/2\pm\sqrt{(\Delta-\Omega_1/2)^2+2\Omega_0^2}\right)/2$ and 
$\omega_b=-\Delta-\Omega_1/2$, respectively. The dressed states of Eq.~(\ref{eq:DressedStates1}) are shown in 
Fig.~\ref{fig:SpinLevels1} (b), along with the oscillator phonons of energy $\omega_{m}$ which couple to the NV center 
via the terms in the second line of Eq.~(\ref{eq:DressedH1}). The effective detunings $\Delta_{1}$ and $\Delta_{2}$ will be 
derived later in the text. 

We note that in our model the coupling field $\Omega_1$ provides external control of the hybridization and eigenfrequencies 
of the single spin levels. In the eigenbasis of $\mathcal{H}_{\text{NV}}$, the mechanical motion couples to two transitions 
of the eigenstates, which is promising for creating mechanical squeezing because of the implied two-phonon transition. A related scheme has been considered for coherent three-level atoms coupled to a cavity field via a two-photon transition for quantum noise quenching and optical field squeezing \cite{Scully:88}. Finally, we note that the orientation of the magnetic field gradient only adds a phase to the mechanical-spin coupling in the eigenbasis.

\begin{figure}[tbp]
\centering
\includegraphics[width=0.65\columnwidth]{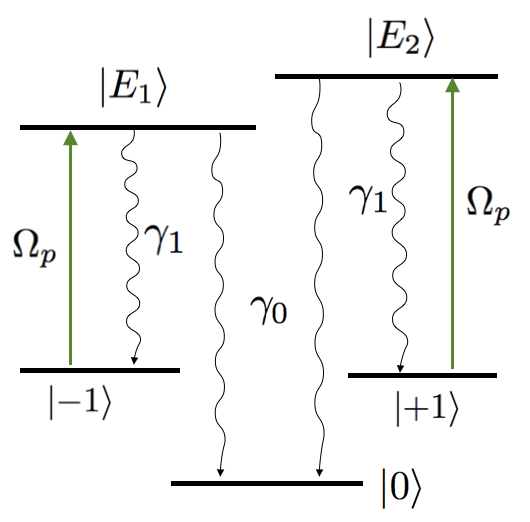}
\caption{ The energy levels of the NV center ground states $\ket{0}$ and $\ket{\pm 1}$ (also shown in Fig. \ref{fig:SpinLevels1} (a)), and excited states $\ket{E_1}$ and $\ket{E_2}$. The optical fields with Rabi frequency $\Omega_p$ pump the population from $\ket{\pm 1}$ to $\ket{E_1}$ and $\ket{E_2}$, which decay to $\ket{0}$ and $\ket{\pm1}$ with effective decay rates $\gamma_0$ and $\gamma_1$, respectively. \label{fig:odissipation}}
\end{figure}

\subsection{Driving-induced dissipation}
The electron spin in the NV center is notable for its long coherence time even at room temperature \cite{Doherty:13,Wrachtrup:06, Bala:09}. In order to induce fast
dissipation in the spin system, which is necessary for generating steady state mechanical squeezing, we apply two optical 
fields with the same Rabi frequency $\Omega_p$ driving the ground-state spin levels to the excited states $\ket{E_1}$ and 
$\ket{E_2}$ via spin-conserving transitions \cite{Manson:06}, which de-excite to states $\ket{\pm1}$ with a decay rate $\gamma_1$, and to the state $\ket{0}$ with an effective decay rate $\gamma_0$, as shown in Fig. \ref{fig:odissipation}. By 
considering spin-mechanical couplings $(g_{1s},g_{1c})$ and microwave fields $\Omega_0$, $\Omega_1$ much weaker than the optical Rabi frequency $\Omega_{p}$, we find 
the steady-state density matrix elements in the dressed eigenbasis of $\ket{a},~\ket{b},~\ket{c}$ due to the dissipation mechanism as
(see Appendix A)
\begin{widetext}
\begin{equation}
\begin{split}
\label{eq:steadyeigen}
\rho_{bb}&=\frac{\left(\Gamma_1-\Gamma_0\right)\Omega_0^2}{\Gamma_0\left(2\Delta-\Omega_1\right)^2+\left(3\Gamma_1+\Gamma_0\right)\Omega_0^2+\Gamma_0\Gamma_{1}^2},\\
\rho_{cc}&=\frac{1}{2}-\frac{1}{2}\rho_{bb}-\frac{1}{2}\frac{\Delta-\Omega_1/2}{\sqrt{(\Delta-\Omega_1/2)^2+2\Omega_0^2}}\frac{\Gamma_0\left(2\Delta-\Omega_1\right)^2+8\Gamma_0\Omega_0^2+\Gamma_0\Gamma_{1}^2}{\Gamma_0\left(2\Delta-\Omega_1\right)^2+\left(3\Gamma_1+\Gamma_0\right)\Omega_0^2+\Gamma_0\Gamma_{1}^2},\\
\rho_{aa}&=\frac{1}{2}-\frac{1}{2}\rho_{bb}+\frac{1}{2}\frac{\Delta-\Omega_1/2}{\sqrt{(\Delta-\Omega_1/2)^2+2\Omega_0^2}}\frac{\Gamma_0\left(2\Delta-\Omega_1\right)^2+8\Gamma_0\Omega_0^2+\Gamma_0\Gamma_{1}^2}{\Gamma_0\left(2\Delta-\Omega_1\right)^2+\left(3\Gamma_1+\Gamma_0\right)\Omega_0^2+\Gamma_0\Gamma_{1}^2},\\
\rho_{ac}&=\frac{\Omega_0/\sqrt{2}}{\sqrt{\left(\Delta-\Omega_1/2\right)^2+2\Omega_0^2}}\frac{\Gamma_0\Gamma_1^2}{\Gamma_0\left(2\Delta-\Omega_1\right)^2+\left(3\Gamma_1+\Gamma_0\right)\Omega_0^2+\Gamma_0\Gamma_{1}^2}-i\frac{\sqrt{2}\Gamma_0\Omega_0\Gamma_{1}}{\Gamma_0\left(2\Delta-\Omega_1\right)^2+\left(3\Gamma_1+\Gamma_0\right)\Omega_0^2+\Gamma_0\Gamma_{1}^2},
\end{split}
\end{equation}
\end{widetext}
where $\Gamma_{0}=\Omega_p^2\gamma_0/(\gamma_1+\gamma_0)^2$, and $\Gamma_{1}=\Omega_p^2/(\gamma_1+\gamma_0)$. As can be seen
from Eq.~(\ref{eq:steadyeigen}), dissipative driving can be used to control the the populations and coherences for the NV
dressed states. However, the mechanical oscillator interacts with the NV spin due to the presence of the magnetic field.
In the steady-state, therefore, the mechanical frequency is shifted by the NV spin, while mechanical motion can be 
engineered via the mechanical-spin interaction through the driving-induced dissipation. We substantiate these statements below.

\subsection{The reduced master equation of the mechanical oscillator}
In the interaction picture, we can write Eq.~(\ref{eq:DressedH1}) as
\begin{equation}
\begin{split}
\mathcal{H}_I=&-\hbar\Delta_0\ket{a}\bra{a}-\hbar\Delta_1\ket{b}\bra{b}\\
&+\hbar\left(g_{1s}\ket{c}\bra{b}d^{\dagger}+g_{1c}d\ket{a}\bra{b}+h.c.\right),
\label{eq:singlemodeHI}
\end{split}
\end{equation}
where $\Delta_0=2\omega_m-\omega_{ac}$, $\Delta_1=\omega_m-\omega_{bc}$, and we have used the rotating-wave approximation. 
The approximation is valid when $(\Delta_0,\Delta_1, \Gamma_0,\Gamma_1) \lesssim \omega_m$. We now trace out the spin degree 
of freedom to obtain the reduced master equation for the mechanical oscillator density matrix $\rho_m$, which is our system of interest i.e.
\begin{equation}
\begin{split}
\label{eq:reducem0}
\dot{\rho}_m&=\text{Tr}_s\left<-\frac{i}{\hbar}[\mathcal{H},\rho]\right>\\
&=-ig_{1s}\left[d^{\dagger},\rho_{bc}\otimes\rho_m\right]-ig_{1c}\left[d,\rho_{ba}\otimes\rho_m\right]+h.c.,
\end{split}
\end{equation}
where $\text{Tr}_s$ denotes the trace over the spin degree of freedom. We note that the steady-state density 
matrix elements $\rho_{ba}$  and $\rho_{bc}$, due to the fast dissipation and strong Rabi frequencies, are zero
to the lowest order. The first-order perturbation of these quantities are given by the spin-mechanical interaction 
$\mathcal{H}_I$ in Appendix B. By substituting for $\rho_{bc}$ and $\rho_{ba}$, we obtain 
\begin{equation}
\begin{split}
\label{eq:reducem1}
\dot{\rho}_m&=A_{-}\mathcal{D}[d]\rho_m+A_{+}\mathcal{D}[d^{\dagger}]\rho_m-i\delta/2[d^{\dagger}d,\rho_m]\\
&+S_1\left(d^{\dagger2}\rho_m-d^{\dagger}\rho_md^{\dagger}\right)/2+S_2\left(\rho_m d^{\dagger2}-d^{\dagger}\rho_md^{\dagger}\right)/2\\
&+S_1^{\ast}\left(\rho_md^2-d\rho_md\right)/2+S_2^{\ast}\left(d^2\rho_m-d\rho_md\right)/2\\
&+\gamma_m (n_{\text{th}}+1)\mathcal{D}[d]\rho_m+\gamma_m n_{\text{th}}\mathcal{D}[d^{\dagger}]\rho_m,
\end{split}
\end{equation}
where $\mathcal{D}[d]\rho_m=\left(2d\rho_m d^{\dagger}-d^{\dagger}d\rho_m-\rho_m d^{\dagger}d\right)/2$ corresponds
to the standard Lindblad operator, and $\gamma_m$ is the effective decay rate of the mechanical oscillator and 
$n_{\text{th}}=1/(e^{\hbar\omega_m/k_BT}-1)$ is the effective mean phonon number due to both the surrounding gas and the trapping beam \cite{Rodenburg:16}. The mechanical fluctuations due to the optical pump $\Omega_{p}$ are negligible as shown in the following discussion. The coefficients in Eq.~(\ref{eq:reducem1}) 
are given by
\begin{equation}
\begin{split}
\delta&=2\text{Im}\left[\frac{|g_{1s}|^2}{M}(k_2\rho_{cc}-k_2\rho_{bb}+k_3\rho_{ca})\right.\\
&\left.+\frac{|g_{1c}|^2}{M}(k_1\rho_{aa}-k_1\rho_{bb}+k_3\rho_{ac})\right],\\
A_{-}&=2\text{Re}\left[\frac{|g_{1s}|^2}{M}(k_2\rho_{cc}+k_3\rho_{ca})+\frac{|g_{1c}|^2}{M}k_1\rho_{bb}\right],\\
A_{+}&=2\text{Re}\left[\frac{|g_{1c}|^2}{M}(k_1\rho_{aa}+k_3\rho_{ac})+\frac{|g_{1s}|^2}{M}k_2\rho_{bb}\right],\\
S_1&=2|g_{1s}g_{1c}|\left(\frac{k_3}{M}\rho_{aa}+\frac{k_3^{\ast}}{M^{\ast}}\rho_{bb}+\frac{k_2}{M}\rho_{ac}\right),\\
S_2&=2|g_{1s}g_{1c}|\left(\frac{k_3^{\ast}}{M^{\ast}}\rho_{cc}+\frac{k_3}{M}\rho_{bb}+\frac{k_1^{\ast}}{M^{\ast}}\rho_{ac}\right).
\end{split}
\end{equation}
In Eq.~\eqref{eq:reducem1}, the terms proportional to $A_{-}$ ($A_{+}$) describe the dissipation-induced cooling 
(heating) due to coupling of the mechanical motion to the transitions from $\ket{c}$ ($\ket{a}$) to $\ket{b}$. The terms 
proportional to $\delta$ are the mechanical frequency shifts due to the mechanical-spin interaction. The terms proportional to 
$S_j$ or $S_j^{\ast}$ denote the mechanical squeezing via a two-phonon transition using the single NV spin. The quantities $k_1, \ k_2, \ k_2$ and $M$ are defined in the Appendix B.

\subsection{System Dynamics - Analytical Results}
\label{subsec:SDAnal}
To study the system dynamics of the mechanical oscillator, we derive from the reduced master equation \eqref{eq:reducem1} that
\begin{eqnarray}
\dot{\braket{d^{\dagger}d}}&=&-(\gamma_m+A_{-}-A_{+})\braket{d^{\dagger}d}+(S_{1}-S_{2})/2\braket{d^{\dagger2}}\nonumber\\
&&+(S_{1}^{\ast}-S_{2}^{\ast})/2\braket{d^{2}}+(\gamma_mn_{\text{th}}+A_{+}),\\
\dot{\braket{d^2}}&=&-(\gamma_m+A_{-}-A_{+}+i\delta)\braket{d^2}\nonumber\\
&&+(S_{1}-S_{2})\braket{d^{\dagger}d}+S_{1}.
\end{eqnarray}
The steady-state solutions to the above equations are given by
\begin{eqnarray}
\braket{d^{\dagger}d}_{\text{ss}}&=&\frac{(\gamma_mn_{\text{th}}+A_{+})+\text{Re}\left[\frac{(S_1^{\ast}-S_2^{\ast})S_1}{\gamma_m+A_{-}
-A_{+}+i\delta}\right]}{(\gamma_m+A_{-}-A_{+})-\text{Re}\left[\frac{|S_1-S_2|^2}{\gamma_m+A_{-}-A_{+}+i\delta}\right]},\label{eq:ddd}\\
\braket{d^2}_{\text{ss}}&=&\frac{(S_1-S_2)\braket{d^{\dagger}d}_{\text{ss}}+S_1}{\gamma_m+A_{-}-A_{+}+i\delta}.\label{eq:d2}
\end{eqnarray}
To obtain maximum mechanical squeezing, we define the quadrature variance rotated in the phase-space plane such that
\begin{equation}
(\Delta x)^2=\frac{1}{4}\left(\braket{d^{\dagger}d}_{\text{ss}}+\braket{d d^{\dagger}}_{\text{ss}}-|\braket{d^2}_{\text{ss}}|-|\braket{d^{\dagger2}}_{\text{ss}}|\right),
\end{equation}
and the criteria for quantum squeezing is given by
\begin{equation}
(\Delta x)^2<\frac{1}{4}.
\end{equation}
We consider $\Delta_1=0$ ($\omega_m=\omega_{bc}$), $\Delta_2=-\Delta-3\Omega_1/2 \ne 0$ ($\omega_m\ne\omega_{ab}$), and $|\Delta_2|\gg (\Gamma_{0}, g)$, 
such that the cooling transition $(\ket{c}\rightarrow \ket{b})$ is resonant and the heating transition $(\ket{a}\rightarrow \ket{c})$ 
is far off-resonant [see Fig. \ref{fig:SpinLevels1} (b)]. Therefore, steady-state squeezing is possible since the spin-mechanical cooling dominates over the spin-mechanical heating. We obtain to the first-order the quantities:
\begin{equation}
\begin{split}
A_{-}&\approx \frac{4|g_{s}^2|\rho_{cc}}{\Gamma_1(1+\sin^2\theta)},\\
S_1&\approx  2|g_{s}g_{c}|\frac{\sin\theta\cos\theta\rho_{aa}}{i\Delta_2(1+\sin^2\theta)}+\frac{4|g_{s}g_{c}|\rho_{ac}}{\Gamma_1(1+\sin^2\theta)},\\
S_2&\approx  2|g_{s}g_{c}|\frac{\sin\theta\cos\theta\rho_{cc}}{-i\Delta_2(1+\sin^2\theta)},\\
\delta&\approx -2\frac{|g^2_{1c}|\rho_{aa}}{\Delta_2}.
\label{eq:coefficients}
\end{split}
\end{equation}
The other quantities $\delta,~ A_{+},~ S_2\approx 0$. Under the condition $\Gamma_0\approx\Gamma_1\ll\omega_m$, required by the 
rotating wave approximation made earlier, we obtain from Eq.~\eqref{eq:steadyeigen} that
\begin{equation}
\begin{split}
\rho_{cc}&\approx \frac{(1+\cos2\theta)^2}{2(1+\cos^22\theta)},\\
\rho_{aa}&\approx \frac{(1-\cos2\theta)^2}{2(1+\cos^22\theta)},\\
\rho_{ac}&\approx \frac{-i\sin2\theta}{\sqrt{1+\cos^22\theta}}\frac{\Gamma_0/2}{\sqrt{2(\Delta-\Omega_1/2)^2+2\Omega_0^2}}.
\end{split}
\end{equation}
The condition $\Gamma_0\approx\Gamma_1$ can be satisfied by applying a strong resonant microwave field coupling the excited spin triplet states \cite{Manson:06, Neumann:09} to suppress the dissipation from $\ket{E_1}$ and $\ket{E_2}$ to $\ket{\pm1}$. Therefore, the steady-state mean phonon number due to the dissipative cooling is given by
\begin{equation}
\braket{d^{\dagger}d}_{\text{ss}}\approx\frac{\gamma_mn_{\text{th}}}{\gamma_m+A_{-}},
\end{equation}
where the cooling rate is given by
\begin{equation}
A_{-}=\frac{g^2}{\Gamma_0}\frac{8\cos^4\theta\sin^2\theta}{(1+\sin^2\theta)(1+\cos^22\theta)},
\end{equation}
which recovers the result in Ref. \cite{Rabl:09} when $\Gamma_0\approx\Gamma_1\ll\omega_m$. %\begin{equation}
%\begin{split}
%S_1&=\frac{-ig_1^2\sin^22\theta}{(1+\sin^2\theta)(1+\cos^22\theta)}\left(\frac{\sin^4\theta}{\Delta_2}+\frac{\sin2\theta}{\sqrt{2}\Omega_0}\right),\\
%S_2&=\frac{ig_1^2\cos^4\theta\sin^22\theta}{\Delta_2(1+\sin^2\theta)(1+\cos^22\theta)},\\
%\delta&=-\frac{4g_1^2\cos^2\theta\sin^4\theta}{\Delta_2(1+\cos^22\theta)}.
%\end{split}
%\end{equation}
By using the above conditions, we obtain

\begin{equation}
\begin{split}
\braket{d^2}_{\text{ss}}&\approx-\frac{S_1}{\gamma_m+A_{-}+i\delta}\left(\braket{d^{\dagger}d}_{\text{ss}}+1\right).
\end{split}
\end{equation}
The quadrature variance is then given by
\begin{equation}
\begin{split}
(\Delta x)^2&\approx\frac{1}{4}\left(1-\left|\frac{2S_1}{\gamma_m+A_{-}+i\delta}\right|\right)\\
&+\frac{1}{2}\left(1-\left|\frac{S_1}{\gamma_m+A_{-}+i\delta}\right|\right)\braket{d^{\dagger}d}_{\text{ss}}.
\label{eq:dx2}
\end{split}
\end{equation}
Using Eq.~(\ref{eq:dx2}), we can see that for
$\left|\frac{S_1}{\gamma_m+A_{-}+i\delta}\right|< 1$, the quadrature $(\Delta x)^2$ can be smaller than $1/4$ when $\braket{d^{\dagger}d}_{\text{ss}}\sim0$, which demonstrates quantum squeezing of the mechanical motion near the ground 
state. We consider numerical parameters explicitly in the next section.

\begin{figure}[tbp]
\centering
\includegraphics[width=0.95\columnwidth]{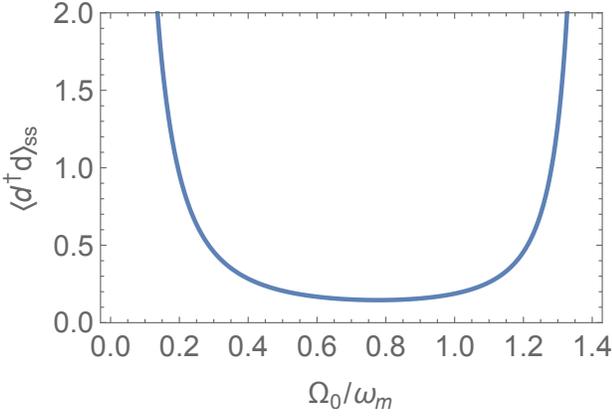}
\caption{Final phonon number versus scaled microwave Rabi frequencies $\Omega_0/\omega_{m}$. The parameters are 
$\omega_m/2\pi=1.0$ MHz, $Q\equiv\omega_m/\gamma_m=10^6$, $n_{\text{th}}=10^3$, $\Omega_1=0$, $\Gamma_{0}=0.25\omega_m$, and $g=0.06\omega_m$.}
\label{fig:twolevel}
\end{figure}

\begin{figure}[tbp]
\centering
\includegraphics[width=0.95\columnwidth]{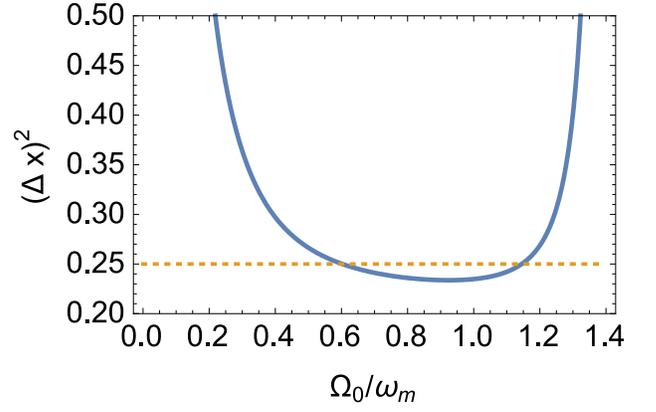}
\caption{Quadrature variance versus $\Omega_0/\omega_{m}$. The parameters are the same as in Fig. \ref{fig:twolevel}.}
\label{fig:dx2case1}
\end{figure}

\begin{figure}[tbp]
\centering
\includegraphics[width=0.95\columnwidth]{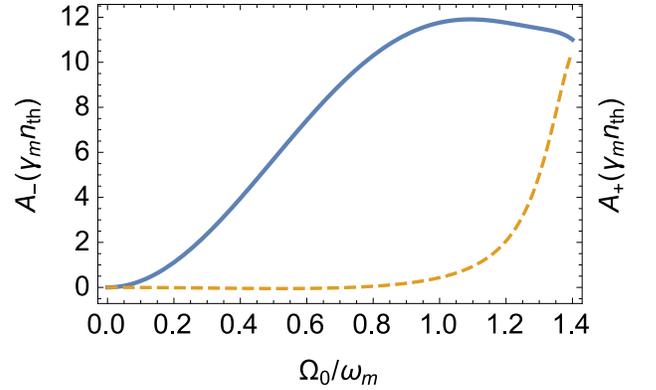}
\caption{Cooling (solid line) and heating (dashed line) rates versus $\Omega_0/\omega_{m}$. The parameters are the same 
as in Fig. \ref{fig:twolevel}.}
\label{fig:coolheatcase1}
\end{figure}

\begin{figure}[tbp]
\centering
\includegraphics[width=0.95\columnwidth]{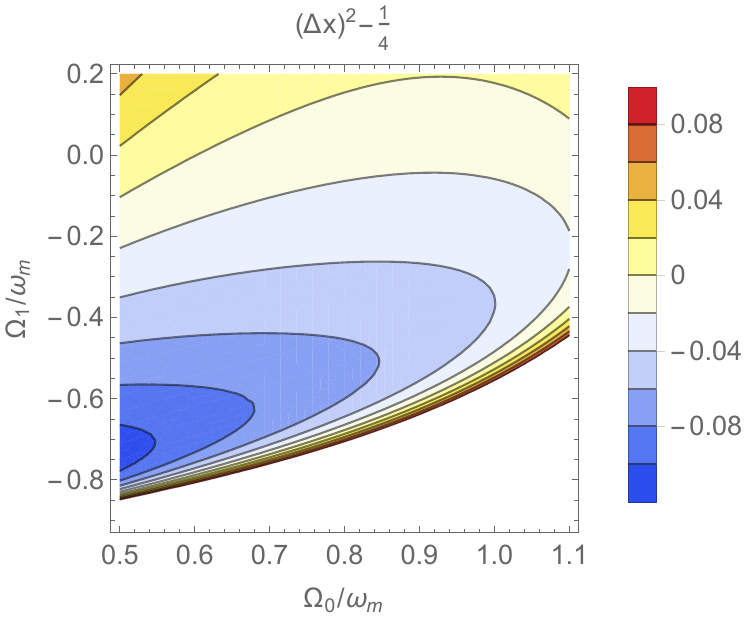}
\caption{Mechanical quadrature squeezing $(\Delta x)^2-1/4$ versus $\Omega_0/\omega_{m}$ and $\Omega_1/\omega_{m}$. 
The other parameters are the same as in Fig. \ref{fig:twolevel}.}\label{fig:dx2vsO}
\end{figure}

\begin{figure}[tbp]
\centering
\includegraphics[width=0.95\columnwidth]{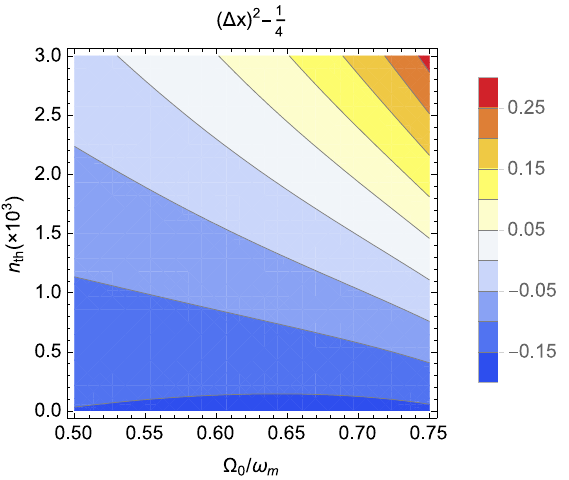}
\caption{Mechanical quadrature squeezing $(\Delta x)^2-1/4$ versus $\Omega_0/\omega_{m}$ and $n_{\text{th}}$ for 
$\Omega_1=-0.7\omega_m$. The other parameters are the same as in Fig. \ref{fig:twolevel}.}\label{fig:dx2vsnth}
\end{figure}

\begin{figure}[tbp]
\centering
\includegraphics[width=0.95\columnwidth]{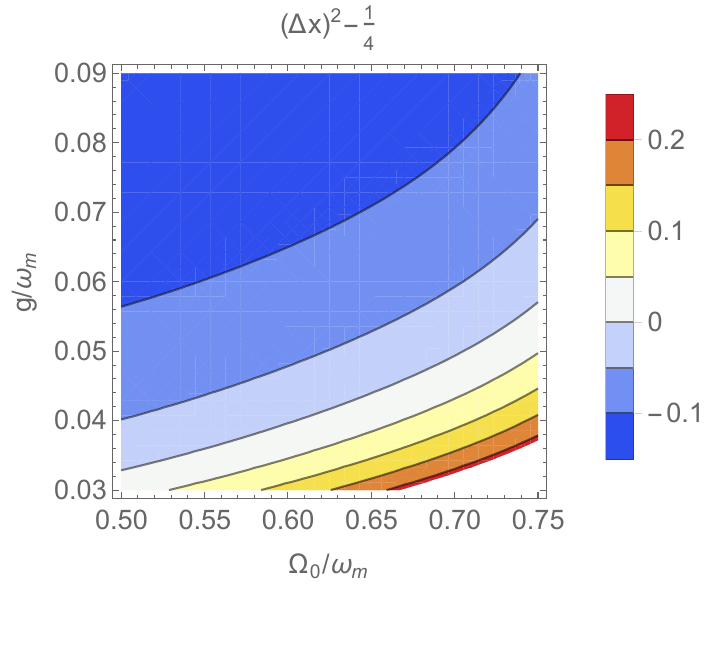}
\caption{Mechanical quadrature squeezing $(\Delta x)^2-1/4$ versus $\Omega_0/\omega_{m}$ and $g$ for 
$\Omega_1=-0.7\omega_m$. The other parameters are the same as in Fig. \ref{fig:twolevel}.}\label{fig:dx2vsOg}
\end{figure}

%\begin{figure}[tbp]
%\centering
%\includegraphics[width=0.95\columnwidth]{dx2vsQ}
%\caption{Mechanical quadrature squeezing ($(\Delta x)^2-1/4$) versus $\Omega_0/\omega_{m}$ and $\log[Q]$ for 
%$\Omega_1=-0.7\omega_m$ at room temperature. The other parameters are the same as in Fig.~\ref{fig:twolevel}.}
%\label{fig:dx2vsQ}
%\end{figure}

\subsection{System Dynamics - Numerical Results}
\subsubsection{Case 1: $\Omega_1=0$}
We first consider the case of no coupling ($\Omega_1=0$) between the $\ket{\pm 1}$ NV states [see 
Fig.~\ref{fig:SpinLevels1}] as this coupling is not essential to the physics, and only provides fine control as shown below. 
We plot the numerical results for $\braket{d^{\dagger}d}_{\text{ss}}$, $(\Delta x)^2$, $A_{-}$, and $A_{+}$ 
using the solutions Eqs. \eqref{eq:ddd} and \eqref{eq:d2}.

First, we observe that ground-state cooling \cite{Rabl:09} is possible with strong cooperativity, i.e.
$g^2/(\Gamma_{0}\gamma_m n_{\text{th}})\gtrsim 1$ as shown in Fig. \ref{fig:twolevel}. In this case the cooling
processes dominate the heating. Second, we observe in
Fig.~\ref{fig:dx2case1} that the quadrature variance $(\Delta x)^2<1/4$, which implies quantum squeezing of the one 
quadrature of the mechanical oscillator. We find that the region for which the quantum squeezing occurs qualitatively 
agrees with the region $\braket{d^{\dagger}d}_{\text{ss}}\ll 1$, as discussed analytically in Section~\ref{subsec:SDAnal}. To 
understand the cooling and the squeezing, we plot $A_{-}$ and $A_{+}$ in Fig. \ref{fig:coolheatcase1}. As 
$\Omega_0/\omega_m$ varies between $0$ and $\sqrt{2}$, we see an optimal cooling limit is obtained by balancing the 
cooling and heating effects from the single spin.

\begin{figure}[tbp]
\centering
\includegraphics[width=0.95\columnwidth]{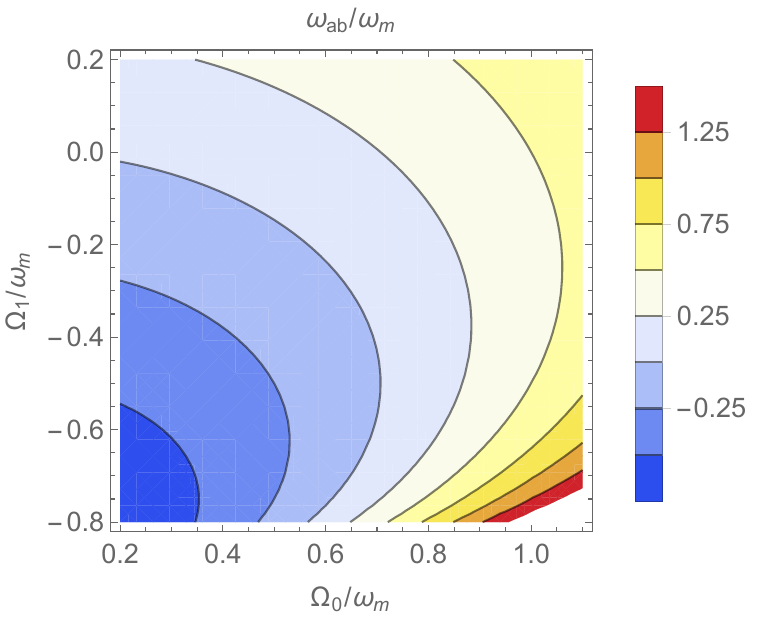}
\caption{The normalized frequency, $\omega_{ab}/\omega_m$, between $\ket{a}$ and $\ket{b}$ versus $\Omega_0$ and $\Omega_1$. The other parameters are the same as in Fig.~\ref{fig:twolevel}.}
\label{fig:3Ddetuning}
\end{figure}

\subsubsection{Case 2: $\Omega_1\ne0$}
For $\Omega_1\ne0$, we have an extra control over the single NV spin which couples to the mechanical oscillator. For an
initial phonon number $n_{\text{th}}=10^3$ and mechanical quality factor $Q=10^6$, we first plot the quantity
$(\Delta x)^2-1/4$ versus the scaled Rabi frequencies $\Omega_0/\omega_{m}$ and $\Omega_1/\omega_{m}$ in 
Fig.~\ref{fig:dx2vsO}. We observe that quantum squeezing, $(\Delta x)^2-1/4<0$, can be realized for a large range of 
parameters. An enhancement of the mechanical squeezing can be obtained for $\Omega_1<0$, which corresponds to $\Omega_1\ne0$ and a $\pi$ phase difference between the driving fields $\Omega_1$ and $\Omega_0$. Second, we plot the quantity $(\Delta x)^2-1/4$ vs $\Omega_0/\omega_{m}$ and $n_{\text{th}}$
in Fig.~\ref{fig:dx2vsnth}, where strong squeezing below $3\text{dB}$ can be obtained, i. e. $(\Delta x)^2-1/4<-1/8$. We 
find quantum squeezing can be achieved when $n_{\text{th}}\sim3\times10^3$, which corresponds to an initial temperature 
$\sim$ $0.1$ K. This initial temperature of the mechanical oscillator may be achieved with cryogenic techniques or by using feedback cooling 
\cite{Gieseler:12,Rodenburg:16, Jain:16}. Furthermore, we plot the quantity $(\Delta x)^2-1/4$ vs $\Omega_0/\omega_{m}$ 
and $g/\omega_{m}$ keeping other parameters constant, in Fig.~\ref{fig:dx2vsOg}. We see from the figure that stronger 
$g$ is preferred for realizing quantum squeezing as long as the Born-Markov approximation is valid.

Remarkably, we find, at initial room temperature environment for the mechanical oscillator, that quantum squeezing is 
feasible with our system for ultrahigh vacuum with feedback cooling. In ultrahigh vacuum ($<10^{-8}$ mbar), as demonstrated recently for an optically levitated nanoparticle \cite{Jain:16}, the gas damping rate is on the order of $\gamma_{g}\sim10^{-6}$ Hz, which corresponds 
to $\omega_m/\gamma_g\sim 10^{12}$. As an example, we consider an optically levitated nanodiamond with a radius $50$ nm and a mechanical oscillation frequency $\omega_m/2\pi=1.0$ MHz along $x$ axis in a magnetic field gradient of $\sim10^5$ T$/$m. Recent experiment has produced a strong magnetic field gradient of $\sim10^6$ T$/$m in a $23$-nm position shift from a magnetic tip \cite{Mamin:07}. To obtain an optical-induced dissipation rate $\Gamma_0=\omega_m/4\approx 1.5$ MHz for the electron spin, we consider an optical pump Rabi frequency $\Omega_p\sim8$ MHz and a typical excited state decay rate $\gamma_0\sim40$ MHz \cite{Doherty:13}. For $\Omega_p\sim8$ MHz, the corresponding optical pump power is smaller than $1$ $\mu$W \cite{Robledo:10}, which has a negligible effect on the mechanical motion fluctuation due to the optical scattering \cite{Chang:10}. To reduce the mean phonon number of the mechanical oscillator due to both the surrounding gas and the optical trapping field, feedback cooling of the nanoparticle can be employed by introducing extra mechanical damping from feedback \cite{Gieseler:12, Rodenburg:16, Jain:16}. We estimate that with a feedback-induced mechanical damping $\gamma_{fb}\sim10^3$ Hz, quantum squeezing is achievable at room temperature when the initial phonon occupation number is reduced to $n_{\text{th}}\sim2$. Our prediction is within the reach of a recent experiment, where a final phonon number of $63$ has been demonstrated with feedback cooling \cite{Jain:16}.

We note that using the driving field $\Omega_1$, it is possible to control the energy difference between dressed states. Our model requires $\omega_{ab}>0$ for the rotating-wave approximation to be valid. We plot the value of $\omega_{ab}$ vs $\Omega_0$ and $\Omega_1$ in Fig.~\ref{fig:3Ddetuning} and we find the condition is satisfied for the parameter regime where mechanical quantum squeezing can be engineered.

To summarize,  single-mode quantum squeezed mechanical state is feasible using our model in ultrahigh vacuum, even at room temperature.

\section{Single NV center coupled to two mechanical modes\label{sec:twomode}}
\begin{figure}[tbp]
\centering
\includegraphics[width=0.95\columnwidth]{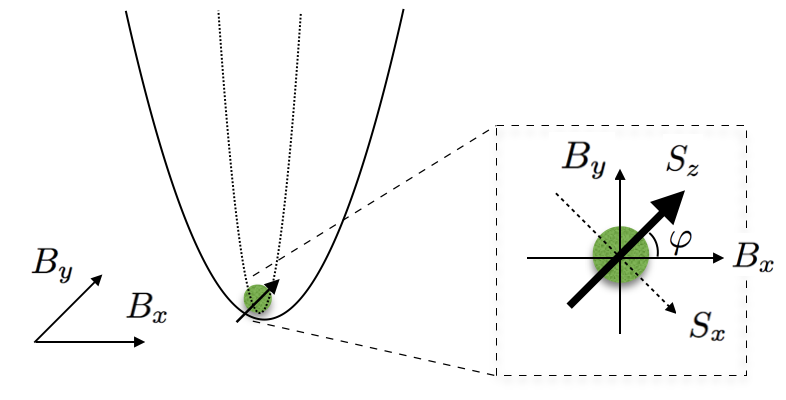}
\caption{The configuration considered in Section \ref{sec:twomode}. The green circle denotes an optically levitated 
nanodiamond oscillating in two separate harmonic potentials along the $x$ coordinate (solid black curve) and $y$ coordinate (dotted black curve), respectively. Magnetic field gradients are applied along both $x$ and $y$ directions.  As shown in the figure, $S_{x},S_{z}$
and $B_{x}, B_{y}$ all lie on the same plane, and the spin axes are not aligned with the coordinate axes. The spin component $S_{y}$ (not shown) is perpendicular to $B_{x}, B_{y}$ and points into the plane. \label{fig:Scheme2}}
\end{figure}
The optically levitated nanodiamond has three harmonic oscillations independent of each other for small oscillation amplitudes, which is an excellent platform for multimode mechanical quantum state engineering. By applying magnetic field gradient in both $x$ and $-y$ directions of the harmonic oscillations, as shown in Fig. \ref{fig:Scheme2}, we can couple two mechanical
modes to the single spin of the NV center nanodiamond. The magnetic field gradient are chosen such that $\nabla\cdot B=0$, i.e., $\frac{\partial B_x}{\partial x}=-\frac{\partial B_y}{\partial y}$. We assume $x$ and $y$ coordinate axes of the mechanical motions are in the
plane of the spin operator components $S_x$ and $S_z$. The angles between $B_x$ and $S_z$, and between $B_y$ and $S_z$, are $\varphi$
and $\pi/2-\varphi$, respectively. The interaction Hamiltonian of the single spin and the mechanical motions are given by
\begin{equation}
\begin{split}
\mathcal{H}_{\text{int}}&=\hbar g_1\cos(\varphi)S_z(d_1^{\dagger}+d_1)+\hbar g_1\sin(\varphi)S_x(d_1^{\dagger}+d_1)\\
&+\hbar g_2\sin(\varphi)S_z(d_2^{\dagger}+d_2)-\hbar g_2\cos(\varphi)S_x(d_2^{\dagger}+d_2),
\end{split}
\end{equation}
where $d_1$ ($d_2$) is the annihilation operator in $x$ ($y$) direction. The spin-mechanical coupling strengths are $g_1=g_l\mu_bB_0 x_0/\hbar$ and $g_2=g_l\mu_bB_0 y_0/\hbar$, respectively. The electron spin dynamics is the same as in the single mechanical mode case, where the spin
is driven by two microwave fields coupling between states $\ket{0}$ and $\ket{\pm1}$, and an effective field coupling between states $\ket{+1}$
and $\ket{-1}$. In the eigenbasis of $\mathcal{H}_{\text{NV}}$, the interaction Hamiltonian is given by
\begin{equation}
\begin{split}
\mathcal{H}&=\hbar\omega_{m_1} d_1^{\dagger}d_1+\hbar\omega_{m_2} d_2^{\dagger}d_2+\hbar\omega_{ac}\ket{a}\bra{a}
+\hbar\omega_{bc}\ket{b}\bra{b}\\
&+\hbar\left(g_{1s}\ket{c}\bra{b}+g_{1c}^{\ast}\ket{b}\bra{a}+h.c.\right)(d_1^{\dagger}+d_1)\\
&+\hbar\left(g_{2s}\ket{c}\bra{b}+g_{2c}^{\ast}\ket{b}\bra{a}+h.c.\right)(d_2^{\dagger}+d_2)\\
&+\hbar \left[g_1\sin(\varphi)(d_1^{\dagger}+d_1)-g_2\cos(\varphi)(d_2^{\dagger}+d_2)\right]\\
&\times\left[\left(\sin^2(\theta)\ket{c}\bra{c}+\cos^2(\theta)\ket{a}\bra{a}-\ket{b}\bra{b}\right)\right.\\
&-\sin(\theta)\cos(\theta)\left(\ket{c}\bra{a}+\ket{a}\bra{c}\right)\big],\label{eq:twomodeH}
\end{split}
\end{equation}
where $g_{1s}=-g_1\cos(\varphi)\sin(\theta)$, $g_{1c}=g_1\cos(\varphi)\cos(\theta)$, $g_{2s}=-g_2\sin(\varphi)\sin(\theta)$, and $g_{2c}=g_2\sin(\varphi)\sin(\theta)$. The other quantities are the same as in the single mechanical mode case. The last term in Eq. \eqref{eq:twomodeH} describes the interaction between the electron spin $S_x$ with the two mechanical modes in the dressed-state basis. We consider the situation that $g_1,g_2\ll\omega_{ab},\omega_{bc}$, therefore the interaction in this term that results in frequency shifts of levels $\ket{a},\ket{b},\ket{c}$ may be neglected. Assuming that $\omega_{m_j}\sim\omega_{ab},\omega_{bc}$, we may also neglect the part proportional to $\left(\ket{c}\bra{a}+\ket{a}\bra{c}\right)$ in the last term since $\omega_{m_j}\lesssim|\omega_{ac}-\omega_{m_j}|$ under the rotating-wave approximation.

By considering the two mode frequencies $\omega_{m_1}=\omega_{m_2}=\omega_m$ resonant coupled to the transition from $\ket{b}$ to $\ket{c}$ and far
detuned from the other transition of $\ket{a}$ to $\ket{b}$, we can write $\mathcal{H}$, in the interaction-picture under the rotating-wave
approximation as
\begin{equation}
\begin{split}
\mathcal{H}_I=&-\hbar\Delta_0\ket{a}\bra{a}-\hbar\Delta_1\ket{b}\bra{b}\\
&+\hbar\left[-|g_{s}|\ket{c}\bra{b}\left(\cos(\varphi)d_1^{\dagger}+\sin(\varphi)d_2^{\dagger}\right)\right.\\
&+|g_{c}|\big(\cos(\varphi)d_1+\sin(\varphi)d_2\big)\ket{a}\bra{b}+h.c.\Big],
\end{split}
\end{equation}
where $\Delta_0=2\omega_m-\omega_{ac}$ and $\Delta_1=\omega_m-\omega_{bc}$ are the same as the single mode case. Similar to the single-mode case, the interaction Hamiltonian for the two-mode is obtained by replacing $d$ with $\cos(\varphi)d_1+\sin(\varphi)d_2$ and $d^{\dagger}$ with $\cos(\varphi)d_1^{\dagger}+\sin(\varphi)d_2^{\dagger}$ in Eq. \eqref{eq:singlemodeHI}.
This configuration is possible for a nanoparticle trapped in an optical field, where the frequencies of two transverse modes can be made very close to each other \cite{Gieseler:12}. The advantage of this configuration is such that the superposed mode $\cos(\varphi)d_1+\sin(\varphi)d_2$ can be cooled efficiently to its ground-state similar to the single mode case while squeezing process is engineered via the two-phonon transition of the superposed mode mediated by the dressed-state spin levels.

At the steady-state of the spin states, we can trace out the spin degree of freedom
to obtain the reduced master equation for the two-mode mechanical oscillator similar to Eq. \eqref{eq:reducem1}
\begin{widetext}
\begin{equation}
\begin{split}
\label{eq:reducem2}
\dot{\rho}_m&=\frac{A_{-}}{2}\mathcal{D}[d_1+d_2]\rho_m+\frac{A_{+}}{2}\mathcal{D}[d_1^{\dagger}+d_2^{\dagger}]\rho_m-i\frac{\delta}{4}[(d_1^{\dagger}+d_2^{\dagger})(d_1+d_2),\rho_m]\\
&+\frac{S_1}{4}\left[(d_1^{\dagger}+d_2^{\dagger})^2\rho_m-(d_1^{\dagger}+d_2^{\dagger})\rho_m(d_1^{\dagger}+d_2^{\dagger})\right]+\frac{S_2}{4}\left[\rho_m(d_1^{\dagger}+d_2^{\dagger})^2-(d_1^{\dagger}+d_2^{\dagger})\rho_m(d_1^{\dagger}+d_2^{\dagger})\right]\\
&+\frac{S_1^{\ast}}{4}\left[\rho_m(d_1+d_2)^2-(d_1+d_2)\rho_m(d_1+d_2)\right]+\frac{S_2^{\ast}}{4}\left[(d_1+d_2)^2\rho_m-(d_1+d_2)\rho_m(d_1+d_2)\right]\\
&+\gamma_m (n_{\text{th}}+1)\mathcal{D}[d_1]\rho_m+\gamma_m n_{\text{th}}\mathcal{D}[d_1^{\dagger}]\rho_m+\gamma_m (n_{\text{th}}+1)\mathcal{D}[d_2]\rho_m+\gamma_m n_{\text{th}}\mathcal{D}[d_2^{\dagger}]\rho_m,
\end{split}
\end{equation}
\end{widetext}
where the coefficients are given in Eq. \eqref{eq:coefficients}, and $\gamma_{m_j}=\gamma_m$ is the decay rate, assumed to the same for both mechanical modes. The angle $\varphi$ is assumed to be $\pi/4$ such that the maximum coupling between both modes can be exploited via the dressed levels. The terms, such as the cooling, the heating, and the squeezing, in the reduced master equation for two mechanical modes are the similar to those of the singe-mode case. We are interested in the steady-state properties of the two-mode system and we find at the steady-state the relevant mean values are
\begin{widetext}
\begin{eqnarray}
\braket{(d_1^{\dagger}+d_2^{\dagger})(d_1+d_2)}_{\text{ss}}&=&\frac{(2\gamma_mn_{\text{th}}+2A_{+})+2\text{Re}\left[\frac{(S_1^{\ast}-S_2^{\ast})S_1}{\gamma_m+A_{-}
-A_{+}+i\delta}\right]}{(\gamma_m+A_{-}-A_{+})-\text{Re}\left[\frac{|S_1-S_2|^2}{\gamma_m+A_{-}-A_{+}+i\delta}\right]},\label{eq:d1dd2d}\\
\braket{(d_1+d_2)^2}_{\text{ss}}&=&\frac{(S_1-S_2)\braket{(d_1^{\dagger}+d_2^{\dagger})(d_1+d_2)}_{\text{ss}}+2S_1}{\gamma_m+A_{-}-A_{+}+i\delta}.\label{eq:d12}
\end{eqnarray}
\end{widetext}
To show two-mode mechanical squeezing, we consider the variance $\braket{\Delta u^2}$ \cite{Gerry:05}, where $u=\left(x_1^{\theta_1}+x_2^{\theta_2}\right)/2$, and $x_j^{\theta_j}=(d_je^{-i\theta_j}+d_j^{\dagger}e^{i\theta_j})/\sqrt{2}$ $(j=1,2)$.
To obtain the maximum degree of two-mode squeezing, we choose $\theta_1$ and $\theta_2$ such that the two-mode quadrature variance is given by
\begin{equation}
\braket{\Delta u^2}=\frac{1}{4}\left(\braket{(d_1^{\dagger}+d_2^{\dagger})(d_1+d_2)}_{\text{ss}}-|\braket{(d_1+d_2)^2}_{\text{ss}}|+1\right).
\end{equation}
We find that in the two-mode case $\braket{(d_1^{\dagger}+d_2^{\dagger})(d_1+d_2)}_{\text{ss}}=2\braket{d^{\dagger}d}_{\text{ss}}$ and $\braket{(d_1+d_2)^2}_{\text{ss}}=\braket{d^2}_{\text{ss}}$ comparing with the single-mode results. Therefore, the  two-mode quadrature variance under current configuration recovers that of the single-mode case, i.e.,
\begin{equation}
\braket{\Delta u^2}=\braket{\Delta x^2}.
\end{equation}
All the discussions about squeezing a single-mode mechanical oscillator apply to the two-mode case exactly under the assumption that the interaction between the spin component $S_x$ and the two mechanical modes are negligible. This assumption we made in the two-mode case is valid in the rotating-wave approximation. We also verify that $\braket{d_j^{\dagger}d_j}_{\text{ss}}>0$ for the parameter regime of interest for the requirement of steady-state of the two modes. 

In summary, we presented a method for engineering two-mode mechanical squeezed states under similar conditions required for the single-mode case. The two-mode squeezed states are controllable over a wide range of parameters even at room temperature and are feasible within current experimental reach, as shown in the single-mode case. As an application, the two-mode mechanical squeezed states are useful for sensitive phase measurement beyond the standard quantum limit in an interferometric setup \cite{Cheung:16, Anisimov:10}.

%\begin{figure}[tbp]
%\begin{subfigure}[t]{0.38\textwidth}
%\centering
%\includegraphics[width=1.0\textwidth]{twomode}
%\end{subfigure}
%\caption{Two-mode squeezing}
%\label{fig:2mode}
%\end{figure}

\section{Conclusion}
In this paper, we have investigated quantum state engineering of an optically levitated nanodiamond coupled to a single NV center ground-state electron spin. We considered quantum state engineering of both single-mode and two-mode mechanical motions. Both analytical and numerical results have been obtained to show that single-mode squeezed states of the mechanical oscillator is feasible with the state-of-art experiments even at room temperature. We have shown that our scheme for single-mode squeezing can be readily extended to the case of two-mode squeezing, which is of interest for precision measurements.

In conclusion, we presented an experimentally realizable method for engineering both single-mode and two-mode mechanical squeezed states in an optically levitated nanodiamond via dressed-state coherence. Our work advances macroscopic quantum state engineering in cavity-free systems, and paves the way for sensitive metrology with squeezed mechanical states.

\section*{ Acknowledgments}
This research is supported by the Office of Naval Research under Award No. N00014-14-1-0803. We thank
A. N. Vamivakas, B. Rodenburg and C. Zou for useful discussions.

\appendix
\section{Driving induced dissipation}
The coupling Hamiltonian for optical pumping is given by $\mathcal{H}_d=\hbar\Omega_p/2\left(\ket{-1}\bra{E_1}+\ket{+1}\bra{E_2}+h.c.\right)$ (see Fig. \ref{fig:odissipation}). We assume effective dissipation paths from the excited states $\ket{E_i}$ to the $\ket{\pm1}$ and $\ket{0}$ with dissipation
rates $\gamma_1$ and $\gamma_0$, respectively. We consider the situation that the driving fields and the decay rates are much
faster than the spin-mechanical coupling such that we can treat the dynamics of the spin separately from the mechanical motion.
The master equation of the spin system is given by $\dot{\rho}=-i/\hbar[\mathcal{H}_d+\mathcal{H}_{\text{NV}},\rho]-1/2\{\Gamma,\rho\}$,
where $\{\Gamma,\rho\}=\Gamma\rho+\rho\Gamma$ with $\Gamma$ the decay matrix of the relevant levels.  By considering
$\Omega_0, \Omega_1\ll\Omega_p, \gamma_1,\gamma_0$, the master equation for the density matrix elements related to the excited levels,
which are dominated by $\mathcal{H}_d$, are given by
\begin{widetext}
\begin{eqnarray}
\dot{\rho}_{E_1E_1}&=&-(\gamma_1+\gamma_0)\rho_{E_1E_1}+i\frac{\Omega_p}{2}\left(\rho_{E_1+1}-\rho_{+1E_1}\right),\\
\dot{\rho}_{E_1+1}&=&-\frac{\gamma_1+\gamma_0}{2}\rho_{E_1+1}-i\frac{\Omega_p}{2}\left(\rho_{+1+1}-\rho_{E_1E_1}\right),\\
\dot{\rho}_{E_2E_2}&=&-(\gamma_1+\gamma_0)\rho_{E_2E_2}+i\frac{\Omega_p}{2}\left(\rho_{E_2-1}-\rho_{-1E_2}\right),\\
\dot{\rho}_{E_2-1}&=&-\frac{\gamma_1+\gamma_0}{2}\rho_{E_2-1}-i\frac{\Omega_p}{2}\left(\rho_{-1-1}-\rho_{E_2E_2}\right),
\label{eq:didis}
\end{eqnarray}
\end{widetext}
At the steady-state, we find from Eqs. (A1)-(A4) that $\rho_{E_1E_1}=\Omega_p^2\rho_{+1+1}/[(\gamma_1+\gamma_0)^2+\Omega_p^2]$, $\rho_{E_2E_2}=\Omega_p^2\rho_{-1-1}/[(\gamma_1+\gamma_0)^2+\Omega_p^2]$, $\rho_{E_1+1}=-i\Omega_p(\gamma_1+\gamma_0)\rho_{+1+1}/[(\gamma_1+\gamma_0)^2+\Omega_p^2]$, and $\rho_{E_2-1}=-i\Omega_p(\gamma_1+\gamma_0)\rho_{-1-1}/[(\gamma_1+\gamma_0)^2+\Omega_p^2]$. We then find the equation of motion
of the density matrix elements for the ground-state spin levels due to the microwave fields as
\begin{widetext}
\begin{eqnarray}
\dot{\rho}_{+1+1}&=&-\Gamma_{0}\rho_{+1+1}-i\frac{\Omega_0}{2}\left(\rho_{0+1}-\rho_{+10}\right)
-i\frac{\Omega_1}{2}\left(\rho_{-1+1}-\rho_{+1-1}\right),\\
\dot{\rho}_{-1-1}&=&-\Gamma_{0}\rho_{-1-1}-i\frac{\Omega_0}{2}\left(\rho_{0-1}-\rho_{-10}\right)
+i\frac{\Omega_1}{2}\left(\rho_{-1+1}-\rho_{+1-1}\right),\\
\dot{\rho}_{00}&=&\Gamma_{0}(\rho_{+1+1}+\rho_{-1-1})+i\frac{\Omega_0}{2}\left(\rho_{0+1}-\rho_{+10}\right)
+i\frac{\Omega_0}{2}\left(\rho_{0-1}-\rho_{-10}\right),\\
\dot{\rho}_{+10}&=&\left(i\Delta-\frac{\Gamma_{1}}{2}\right)\rho_{+10}-i\frac{\Omega_0}{2}\left(\rho_{00}
-\rho_{+1+1}\right)+i\frac{\Omega_0}{2}\rho_{+1-1}-i\frac{\Omega_1}{2}\rho_{-10},\\
\dot{\rho}_{-10}&=&\left(i\Delta-\frac{\Gamma_{1}}{2}\right)\rho_{-10}-i\frac{\Omega_0}{2}\left(\rho_{00}
-\rho_{-1-1}\right)+i\frac{\Omega_0}{2}\rho_{-1+1}-i\frac{\Omega_1}{2}\rho_{+10},\\
\dot{\rho}_{-1+1}&=&-\Gamma_{1}\rho_{-1+1}-i\frac{\Omega_0}{2}\left(\rho_{0+1}-\rho_{-10}\right)
-i\frac{\Omega_1}{2}\left(\rho_{+1+1}-\rho_{-1-1}\right),
\end{eqnarray}
\end{widetext}
where $\Gamma_{0}=\Omega_p^2\gamma_0/[(\gamma_1+\gamma_0)^2+\Omega_p^2]\approx\Omega_p^2\gamma_0/(\gamma_1+\gamma_0)^2$, and $\Gamma_{1}=\Omega_p^2/(\gamma_1+\gamma_0)$. We find the steady-state solutions to Eqs. (A5)--(A10) as
\begin{widetext}
\begin{equation}
\begin{split}
\rho_{00}&=\frac{\Gamma_0\left(2\Delta-\Omega_1\right)^2+\left(\Gamma_1+\Gamma_0\right)\Omega_0^2
+\Gamma_0\Gamma_{1}^2}{\Gamma_0\left(2\Delta-\Omega_1\right)^2+\left(3\Gamma_1+\Gamma_0\right)\Omega_0^2+\Gamma_0\Gamma_{1}^2},\
\rho_{+1+1}=\frac{\Gamma_1\Omega_0^2}{\Gamma_0\left(2\Delta-\Omega_1\right)^2+\left(3\Gamma_1
+\Gamma_0\right)\Omega_0^2+\Gamma_0\Gamma_{1}^2},\\
\rho_{-1-1}&=\frac{\Gamma_1\Omega_0^2}{\Gamma_0\left(2\Delta-\Omega_1\right)^2+\left(3\Gamma_1
+\Gamma_0\right)\Omega_0^2+\Gamma_0\Gamma_{1}^2},\
\rho_{-1+1}=\frac{\Gamma_0\Omega_0^2}{\Gamma_0\left(2\Delta-\Omega_1\right)^2+\left(3\Gamma_1
+\Gamma_0\right)\Omega_0^2+\Gamma_0\Gamma_{1}^2},\\
\rho_{-10}&=\frac{\Gamma_0\Omega_0\left(2\Delta-\Omega_1-i\Gamma_{1}\right)}
{\Gamma_0\left(2\Delta-\Omega_1\right)^2+\left(3\Gamma_1+\Gamma_0\right)\Omega_0^2+\Gamma_0\Gamma_{1}^2},\
\rho_{+10}=\frac{\Gamma_0\Omega_0\left(2\Delta-\Omega_1-i\Gamma_{1}\right)}
{\Gamma_0\left(2\Delta-\Omega_1\right)^2+\left(3\Gamma_1+\Gamma_0\right)\Omega_0^2+\Gamma_0\Gamma_{1}^2}.
\end{split}
\end{equation}
\end{widetext}
The steady-state solutions can be rearranged to give the results in Eq. \eqref{eq:steadyeigen} in the eigenbasis.
\section{Reduced master equation}
The first-order perturbation of $\rho_{bc}$ and $\rho_{ba}$ are given by the spin-mechanical interaction $\mathcal{H}_I$ as
\begin{widetext}
\begin{eqnarray}
\dot{\rho}_{bc}\otimes\rho_m&\approx&\left(i\Delta_1-\frac{\Gamma_{1}}{2}(1+\sin^2\theta)\right)\rho_{bc}
\otimes\rho_m+\frac{\Gamma_{1}}{2}\sin(\theta)\cos(\theta)\rho_{ba}\otimes\rho_m\nonumber\\
&&-ig_{s}^{\ast}\left(d\rho_{cc}\otimes\rho_m-\rho_{bb}\otimes\rho_m d\right)-ig_{c}^{\ast}d^{\dagger}
\rho_{ac}\otimes\rho_m,\\
\dot{\rho}_{ba}\otimes\rho_m&\approx&\left(-i\Delta_2-\frac{\Gamma_{1}}{2}(1+\cos^2\theta)\right)\rho_{ba}
\otimes\rho_m+\frac{\Gamma_{1}}{2}\sin(\theta)\cos(\theta)\rho_{bc}\otimes\rho_m\nonumber\\
&&-ig_{c}^{\ast}\left(d^{\dagger}\rho_{aa}\otimes\rho_m-\rho_{bb}\otimes\rho_md^{\dagger}\right)
-ig_{s}^{\ast}d\rho_{ca}\otimes\rho_m,
\end{eqnarray}
\end{widetext}
where $\Delta_2=\Delta_0-\Delta_1=\omega_m-\omega_{ab}$, and the approximation is made on the decay rates of
$\rho_{bc}$ and $\rho_{ba}$ by assuming $\Gamma_0\approx\Gamma_1$. For $|\Delta_j|, \Gamma_{0}\gg g_{s},g_{c}$,
at the steady-state we find
\begin{widetext}
\begin{eqnarray}
\rho_{bc}\otimes\rho_m&=&-ig_{s}^{\ast}\left[\frac{k_2}{M}\left(d\rho_{cc}\otimes\rho_m-\rho_{bb}\otimes\rho_md\right)
+\frac{k_3}{M}d\rho_{ca}\otimes\rho_m\right]\nonumber\\
&&-ig_{c}^{\ast}\left[\frac{k_3}{M}\left(d^{\dagger}\rho_{aa}\otimes\rho_m-\rho_{bb}\otimes\rho_md^{\dagger}\right)
+\frac{k_2}{M}d^{\dagger}\rho_{ac}\otimes\rho_m\right],\\
\rho_{ba}\otimes\rho_m&=&-ig_{c}^{\ast}\left[\frac{k_1}{M}\left(d^{\dagger}\rho_{aa}\otimes\rho_m
-\rho_{bb}\otimes\rho_md^{\dagger}\right)+\frac{k_3}{M}d^{\dagger}\rho_{ac}\otimes\rho_m\right]\nonumber\\
&&-ig_{s}^{\ast}\left[\frac{k_3}{M}\left(d\rho_{cc}\otimes\rho_m-\rho_{bb}\otimes\rho_md\right)+
\frac{k_1}{M}d\rho_{ca}\otimes\rho_m\right],
\end{eqnarray}
\end{widetext}
where $M=k_1k_2-k_3^2$, $k_1=-i\Delta_1+\frac{\Gamma_{1}}{2}(1+\sin^2\theta)$, $k_2=i\Delta_2
+\frac{\Gamma_{1}}{2}(1+\cos^2\theta)$, and $k_3=\frac{\Gamma_{1}}{2}\sin(\theta)\cos(\theta)$. By substituting $\rho_{bc}$ and $\rho_{ba}$ in Eq. \eqref{eq:reducem0}, we obtain the result of the reduced master equation in Eq. \eqref{eq:reducem1}.

\end{document}